\documentclass[twocolumn]{fairmeta}

\usepackage{pifont}
\usepackage{enumitem}
\usepackage{amsmath}
\usepackage{amssymb}
\usepackage{subcaption}

\AtBeginDocument{%
  }

\begin{document}

\title{LLaTTE: Scaling Laws for Multi-Stage Sequence Modeling in Large-Scale Ads Recommendation}

\author[\dagger]{Lee Xiong}
\author[\dagger]{Zhirong Chen}
\author[\dagger]{Rahul Mayuranath}
\author{Shangran Qiu}
\author{Arda Ozdemir}
\author{Lu Li}
\author{Yang Hu}
\author{Dave Li}
\author{Jingtao Ren}
\author{Howard Cheng}
\author{Fabian Souto Herrera}
\author{Ahmed Agiza}
\author{Allen Lin}
\author{Baruch Epshtein}
\author{Anuj Aggarwal}
\author{Julia Ulziisaikhan}

\author{Chao Wang}
\author{Dinesh Ramasamy}
\author{Parshva Doshi}
\author{Sri Reddy}
\author{Arnold Overwijk}

\affiliation[\dagger]{First Authors}
\affiliation{AI at Meta}


\abstract{
We present \textbf{LLaTTE} (LLM-Style Latent Transformers for Temporal Events), a scalable transformer architecture for production ads recommendation. Through systematic experiments, we demonstrate that sequence modeling in recommendation systems follows predictable power-law scaling similar to LLMs. Crucially, we find that semantic features \textit{bend the scaling curve}: they are a prerequisite for scaling, enabling the model to effectively utilize the capacity of deeper and longer architectures. To realize the benefits of continued scaling under strict latency constraints, we introduce a two-stage architecture that offloads the heavy computation of large, long-context models to an asynchronous upstream user model. We demonstrate that upstream improvements transfer predictably to downstream ranking tasks. Deployed as the largest user model at Meta, this multi-stage framework \textbf{drives a 4.3\% conversion uplift on Facebook Feed and Reels} with minimal serving overhead, establishing a practical blueprint for harnessing scaling laws in industrial recommender systems.
}



\metadata[Keywords]{
Recommender Systems, Sequential Modeling, Transformers, Scaling Laws, Multi-stage Ranking
}

\maketitle

\section{Introduction}

The convergence of sequence modeling and recommendation systems has opened new frontiers in artificial intelligence. Modern recommendation systems, particularly in advertising and content platforms, process billions of user interactions daily as each user's journey through products, ads, and content forms a rich temporal sequence that encodes user preferences and intent. While Large Language Models (LLMs) have demonstrated that deep sequence models can achieve remarkable performance through predictable scaling laws~\cite{kaplan2020scaling} (with capabilities improving as power-law functions of model size, data, and compute), production recommendation systems have largely remained constrained to shallower sequence architectures~\cite{kang2018sasrec,sun2019bert4rec,chen2019bst}. 

This gap persists despite the sequential nature of user behavior and the potential for similar scaling benefits. The challenges are twofold: (i) reconciling the computational demands of deep sequence models with the strict latency requirements of production serving, where systems must rank hundreds of candidates within milliseconds, and (ii) bridging the architectural divide between traditional Factorization Machine (FM) based models~\cite{wukong2024} and transformer-based sequence models. FM-based models excel at learning from massive sparse and dense feature spaces (user IDs, item IDs, request and contextual features), but they lack sequential modeling capacity, while sequence models capture temporal dynamics but struggle to efficiently incorporate the high-dimensional non-sequence features critical for production performance. Understanding how to effectively scale sequence models while preserving the strengths of both paradigms under production constraints, and harness predictable scaling laws in this hybrid regime, remains a challenge that bridges research and real-world deployment.

To address these challenges, we structured our investigation around three research questions: \textbf{RQ1}, How can we design sequence models that integrate with FM-based architectures to leverage both sparse collaborative signals and temporal dynamics? \textbf{RQ2}, Do recommendation systems exhibit predictable scaling behaviors similar to LLMs, with identifiable trade-offs, capacity bottlenecks, and data quality impacts? \textbf{RQ3}, How can we harness scaling benefits while meeting strict production latency requirements?

We introduce \textbf{LLaTTE} (LLM-Style Latent Transformers for Temporal Events), an architectural paradigm deployed in our massive-scale production systems that addresses all three questions. LLaTTE enables efficient scaling through two key innovations that reconcile deep sequence modeling with our production constraints:

\begin{enumerate}[leftmargin=*,noitemsep]

    \item \textbf{Target-Aware Adaptive Transformer}: Our sequence module integrates non-sequence sparse features and candidate information into extended query tokens, leveraging Multi-head Latent Attention (MLA)~\cite{deepseek2024mla}. The architecture supports adaptive pyramidal output extraction, which can optionally be employed to progressively reduce computational complexity. This design seamlessly integrates with FM-style modules, enabling us to harness the scaling gains of deep sequence processing within strict production inference budgets.

    \item \textbf{Multi-Stage Architecture}: To harvest scaling benefits under strict latency constraints, we introduce an \textbf{upstream} stage that runs large LLaTTE encoders, triggered by high-value user events, to asynchronously generate and cache compressed user embeddings that are not bounded by request-time latency. The \textbf{online} ranking stage combines these cached embeddings with fresh, short-horizon sequential signals processed by a lightweight LLaTTE counterpart. Both stages share a common sequence architecture while operating at drastically different scales. Specifically, the upstream model consumes $>45\times$ the sequence FLOPs of its online counterpart. This asymmetry allows us to offload the vast majority of sequence computation without impacting online serving latency.

\end{enumerate}

Through systematic experiments spanning model depth, width, sequence length, data enrichment via content understanding model features, and cross-stage transfer dynamics, we provide comprehensive answers to foundational scaling behavior questions in the recommendation system setting, revealing several key insights: (i) Performance follows predictable log-linear scaling laws, with sequence length serving as a primary lever; (ii) Data enrichment such as semantic content embeddings is not merely additive but a \textit{prerequisite} for steeper scaling curves, effectively bending the scaling laws beyond what sparse ID signals allow; (iii) Improvements from upstream modeling transfer predictably to online ranking stages with a high transfer ratio ($\approx 50\%$), validating the efficacy of our multi-stage design even across a strict information bottleneck; and (iv) Model width acts as a capacity bottleneck; sufficient width must be established before depth scaling becomes effective. These findings establish that recommendation systems can benefit from scaling laws analogous to LLMs when architectural constraints are properly addressed.

Our production deployment validates these findings, demonstrating that the two-stage architecture achieves a \textbf{0.25\%} improvement in Normalized Entropy on our primary revenue-generating models with minimal serving overhead. Together, our architectural innovations and empirical scaling laws provide a systematic framework for scaling sequence models in large-scale production recommendation systems.
\section{Background and Related Work}
\label{sec:related}

Our work intersects three research areas: transformer-based sequence modeling in recommendation systems, scaling laws for neural networks, and multi-stage architectures for production deployment. We review each area and position our contributions.

\subsection{Sequence Modeling in Recommendation}

Early sequential recommendation approaches used 
recurrent neural networks for modeling sequential behavior. GRU4Rec~\cite{hidasi2016gru4rec} pioneered RNNs for session-based recommendation, followed by various LSTM variants~\cite{donkers2017sequential}. The introduction of self-attention mechanisms marked a significant advance in the field. SASRec~\cite{kang2018sasrec} and HLLM~\cite{hllm} applied self-attention to next-item prediction, enabling parallel processing and better long-range modeling. BERT4Rec~\cite{sun2019bert4rec} adapted bidirectional transformers using masked item prediction, while S3-Rec~\cite{zhou2020s3rec} introduced self-supervised pretraining for sequential recommendation.

Target-aware attention mechanisms have also proven critical for production CTR prediction. DIN~\cite{zhou2018deep} introduced attention weighted by the target item, later extended by DIEN~\cite{zhou2019deep} with interest evolution modeling.

Recent work has explored different architectural choices for integrating sequential and non-sequential features. HSTU~\cite{hstu}, OneTrans~\cite{onetrans2025}, and LONGER~\cite{longer_2025} advocate for pure sequence-based architectures, while InterFormer~\cite{interformer} proposes interleaving sequence and non-sequence computation at each layer. However, these approaches require tight coupling to specific architectures, limiting flexibility for independent scaling of sequence modules and preventing easy separation of model inference across online and offline stages. This architectural rigidity has constrained production systems to relatively small sequence models processing limited context.

\subsection{Scaling Laws}

Kaplan et al.~\cite{kaplan2020scaling} demonstrated that language model performance follows power-law relationships with model size, dataset size, and compute budget. While recent work has explored scaling in recommender systems~\cite{onetrans2025, wukong2024}, these studies typically examine either individual scaling dimensions in isolation or their compounding effects, without systematically investigating how different scaling axes interact. Moreover, they overlook critical dimensions such as data richness and composition, which are particularly important for large-scale production recommendation systems. We address this by modeling the joint interaction of depth, width, and feature richness, identifying critical thresholds where scaling behavior fundamentally shifts.

\subsection{Multi-Stage Architectures in Production RecSys}

Multi-stage architectures, which separate representation learning from task-specific prediction, have emerged as a practical paradigm for modeling in production recommendation systems. These approaches aim to balance model expressiveness with the stringent latency requirements of online serving.

Embedding-based approaches precompute user representations for consumption by online ranking models. Pinnerformer~\cite{pinnerformer} learns user embeddings from sequential engagement data using a transformer trained with a dense all-action loss over engagement windows. SUM~\cite{sum_user_model_24} provides a framework for sharing user representations across hundreds of production models via asynchronous inference. Both systems decouple upstream computation from online serving, enabling larger models than real-time latency budgets would permit.

TransAct~\cite{transact} introduces a hybrid architecture that combines precomputed embeddings with real-time processing of recent user actions. The system processes recent actions ($O(10^2)$) through a transformer while leveraging precomputed embeddings for long-term interests. TransAct V2~\cite{transact_v2} extends the real-time component to longer sequences ($O(10^4)$) using candidate-anchored nearest neighbor search and custom GPU kernels.

An alternative approach uses retrieval-based designs that operate synchronously within each request. SIM~\cite{pi2020sim} introduced the GSU-ESU paradigm, where a General Search Unit retrieves relevant behaviors before an Exact Search Unit applies target-aware attention. TWIN~\cite{twin2023} improves consistency between stages by ensuring both use identical relevance metrics, scaling to $10^4$--$10^5$ behaviors at Kuaishou. TWIN V2~\cite{twinv2024} extends to lifecycle-scale sequences through hierarchical clustering compression.

Despite these advances, prior multi-stage work has not explored systematic model scaling following predictable scaling laws. These systems typically scale along a single dimension (e.g., sequence length in TransAct V2) while keeping models relatively small overall. In this work, we show that a unified scaling framework applies to both real-time and asynchronous upstream ranking settings. We systematically examine how increasing model capacity across multiple dimensions and allocating capacity between online and upstream stages, affects performance in production recommendation systems. To our knowledge, this is the first work to establish empirical scaling laws for such hybrid upstream-downstream architectures. Additionally, we rigorously quantify the \textit{transfer ratio} between stages, offering a metric to measure how effectively scaling laws traverse the inference bottleneck.

\section{Preliminaries: Ads Recommendation Stack}
\label{subsec:formulation}

We consider a standard multi-task ads ranking problem where the model predicts engagement probabilities such as click-through rate (CTR) and conversion rate (CVR). These calibrated probabilities are utilized for both downstream ranking and auction pricing.

\paragraph{Task}
Given a user $u$, candidate ad $i$, and request context $c$, the model predicts a vector of engagement probabilities:
\begin{equation}
\mathbf{y} = f(\mathbf{x}_u, \mathbf{x}_i, \mathbf{x}_{ui}, \mathbf{x}_c; \boldsymbol{\theta}) \in [0,1]^{|\mathcal{H}|},
\end{equation}
where $\mathcal{H} = \{\text{CTR}, \text{CVR}, \ldots\}$ is the set of prediction heads and $\boldsymbol{\theta}$ denotes model parameters.

\paragraph{Feature Structure}
We categorize input features along two axes. By \textbf{source}, we distinguish: (1) \textbf{User} $\mathbf{x}_u$ (demographics, aggregates, behavioral sequence $S_u$); (2) \textbf{Ad} $\mathbf{x}_i$ (creative attributes, landing page, advertiser metadata); (3) \textbf{User-Ad} $\mathbf{x}_{ui}$ (historical interactions); and (4) \textbf{Context} $\mathbf{x}_c$ (time, device, surface).

By \textbf{type}, features are processed as:
\begin{itemize}[leftmargin=*,noitemsep]
    \item \textbf{Sparse IDs} $\mathbf{x}_{\text{sparse}} \in \mathcal{V}^{m_{\text{sparse}}}$: categorical entities mapped to embeddings.
    \item \textbf{Dense features} $\mathbf{x}_{\text{dense}} \in \mathbb{R}^{m_{\text{dense}}}$: precomputed embeddings (e.g., from content encoders).
    \item \textbf{Float attributes} $\mathbf{x}_{\text{float}} \in \mathbb{R}^{m_{\text{float}}}$: continuous variables.
    \item \textbf{Sequences} $S_u$: ordered lists of temporal events.
\end{itemize}

\paragraph{User Sequences}
User behavioral sequence $S_u = \{a_1, a_2, \ldots, a_T\}$ is a temporally ordered list of actions. Each action $a_t$ 
consists of a timestamp, action type, item identifier, surface type, optional content embeddings and other metadata. In our experiments, sequence lengths $T$ range from $500$ to $5000$ actions.

\paragraph{Objective and Evaluation}
We train using a weighted multi-task binary cross-entropy loss. Our primary evaluation metric is the relative improvement in \emph{Normalized Entropy} (NE). NE is defined as the average log loss normalized by the entropy of the empirical CTR:
\begin{equation}
\label{eq:normalized_entropy}
\text{NE} = \frac{-\frac{1}{N}\sum_{i=1}^N \big[y_i \log \hat{y}_i + (1-y_i)\log(1-\hat{y}_i)\big]}{-\big[p \log p + (1-p)\log(1-p)\big]},
\end{equation}
where $p$ is the empirical positive rate observed from the ground truth. NE is a standard metric in industrial recommendation. 

\section{Experimental Backbone: The LLaTTE Paradigm}
\label{sec:llatte}

\begin{figure*}[t]
  \centering
  \includegraphics[width=\textwidth]{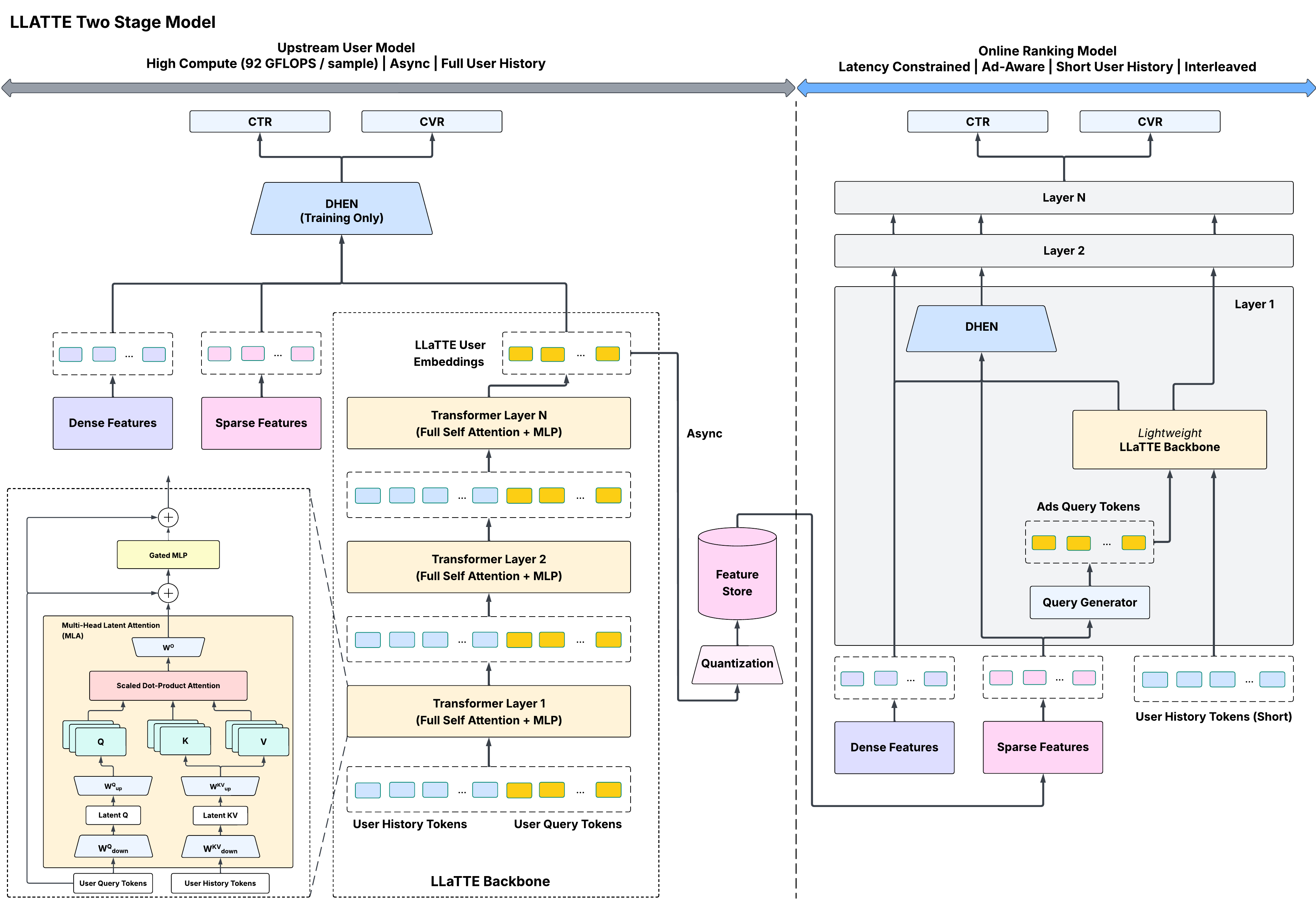}
  \caption{LLaTTE architecture overview. We hold the non-sequence backbone (DHEN) and task heads fixed, and scale the transformer-based sequence module in depth, width, and sequence length. The sequence module utilizes Multi-head Latent Attention (MLA) and adaptive pyramidal trimming to efficiently process long user histories.}
  \label{fig:llatte-overview}
\end{figure*}

To systematically study scaling laws in recommendation, we require an architecture that is both representative of modern production systems and efficient enough to scale to massive contexts. We adopt a modular design, referred to as \textbf{LLaTTE} (LLM-Style Latent Transformers for Temporal Events), that synthesizes best practices in efficient sequence modeling.

The architecture (Figure~\ref{fig:llatte-overview}) consists of three components:
(i) a \textbf{Sequence Module} that encodes user behavioral histories;
(ii) a \textbf{Non-Sequence Module} that integrates sequence summaries with static features; and
(iii) \textbf{Task Heads} for prediction.

\textbf{Compute Allocation.} In our production setting, the sequence module accounts for over 90\% of total inference FLOPs in the large-scale \textbf{upstream} model, while a lightweight sequence module counterpart accounts for roughly 30\% of the \textbf{online} ranking model FLOPs. In subsequent scaling experiments, we focus our study on the large-scale upstream model by default, where varying the capacity ($L, d, T$) of the Sequence Module allows us to explore scaling behaviors across a much wider compute FLOP range than is possible with the lightweight online ranking model. The Non-Sequence backbone and Task Heads are held fixed throughout to isolate the effects of Sequence Module scaling.

\subsection{Non-Sequence Backbone}
\label{subsec:nonseq}

The non-sequence module implements a standard DHEN-style architecture~\cite{zhang2022dhen}, a deep ensemble network that combines heterogeneous feature interaction modules. It processes all non-sequential features alongside the sequence summaries $\mathbf{z}^{(k)}_{\text{seq}}$ produced by the transformer.

Sparse categorical features are embedded and concatenated with dense/float features. The concatenated features are passed through $L_{\text{NS}}$ layers of feature interaction networks to yield a unified representation $\mathbf{z}$. Prediction heads are shallow MLPs producing probabilities $\hat{y}_h = \sigma(\text{MLP}_h(\mathbf{z}))$.

\subsection{Sequence Module}
\label{subsec:seq-overview}

The sequence module is the focus of our scaling study. To enable scaling to long contexts (lengths $>1000$) under production constraints, we employ a \textbf{Target-Aware} transformer with two key efficiency optimizations: \textbf{Multi-head Latent Attention (MLA)} and \textbf{Adaptive Pyramidal Output}.

The module operates in five steps:
\begin{enumerate}[leftmargin=*,noitemsep]
    \item \textbf{Tokenization:} Embeds each action $a_t$ into a token vector $\mathbf{x}_t$, forming $\mathbf{X}_{\text{seq}}$.
    \item \textbf{Fusion:} Appends $n_q$ query tokens to form $\mathbf{X}_{\text{input}}$. We encode the query tokens with user-ad candidate context for online model, and user-only context for upstream model when ad candidate context isn't available yet. This flexible encoding allows the same architecture to bridge offline and online stages.
    \item \textbf{Latent Attention:} Applies an $L$-layer transformer using MLA~\cite{deepseek2024mla} to reduce memory footprint.
    \item \textbf{Pyramidal Reduction:} Selectively trims older tokens at deeper layers (following~\cite{onetrans2025}). This concentrates compute on recent events and query tokens, allowing the model to trade off context length against FLOPs.
    \item \textbf{Readout:} Extracts fixed-size summaries from the final query-token representations.
\end{enumerate}

This adaptive design allows us to deploy different flavors of LLaTTE across the stack while maintaining a unified modeling paradigm. In the asynchronous \textbf{upstream} stage, we deploy high-capacity variants that consume \textbf{$>45\times$ the sequence FLOPs} of the main ranker (with $>90\%$ concentrated in the Sequence Module), utilizing full self-attention to maximize representation quality. Conversely, \textbf{online} ranking models employ aggressive pyramidal trimming to adhere to strict latency budgets. The architecture composes standardized components; full specifications appear in Appendix~\ref{app:llatte-arch}.
\section{Scaling Framework and Methodology}
\label{sec:scaling_framework}

To rigorously quantify the benefits of scaling in recommendation, we move beyond ad-hoc architecture search and adopt a systematic scaling framework. Unlike LLMs, where the input text distribution is generally treated as fixed, recommendation systems allow us to scale along both architectural dimensions and input information densities.

We formulate the scaling behavior as a function of the compute budget $\mathcal{C}$, measured in FLOPs allocated to the sequence module. When the information density is held constant, we posit that Normalized Entropy (NE) improvements (relative to our strong production baseline) follow a power-law relationship with compute, modeled log-linearly as:
\begin{equation}
\label{eq:scaling_law}
\Delta \mathrm{NE}(\mathcal{C}) \propto - \alpha \cdot \log_{10} \mathcal{C}
\end{equation}
where $\alpha$ represents the scaling coefficient (efficiency) of a specific dimension. We investigate four primary axes of scaling:

\paragraph{1. Model Capacity ($L, d$).}
We vary the transformer depth $L$ and embedding width $d$. A key objective is to identify the \textit{aspect ratio} $L/d$ that maximizes parameter efficiency, determining if critical width thresholds exist before depth scaling becomes effective in the sparse-feature regime of recommendation systems.

\paragraph{2. Temporal Horizon ($T$).}
We treat sequence length $T$ as a direct proxy for the information horizon. We examine if extending $T$ yields diminishing returns or if larger models can effectively attend to increasingly distant history events ($T \to 5000$).

\paragraph{3. Information Density.}
Standard scaling laws assume a homogeneous token distribution. We challenge this by varying the semantic richness of the tokens $\mathbf{x}_t$. We compare sparse ID-based tokens against tokens enriched with dense semantic embeddings from foundational content encoders, treating \textit{signal quality} as a multiplier on the scaling coefficient $\alpha$.

\paragraph{4. Cross-Stage Transfer Efficiency.}

To quantify cross-stage efficiency, we introduce the \textbf{Transfer Ratio} ($\tau$) which measures how effectively improvements from a large upstream model translate to the production online ranking loss. This metric accounts for all production interference factors such as asynchronous inference latency penalties.

\begin{equation}
\label{eq:transfer_ratio}
\tau = \frac{\Delta \mathrm{NE}_{\text{downstream}}}{\Delta \mathrm{NE}_{\text{upstream}}}
\end{equation}
where $\Delta \mathrm{NE}_{\text{upstream}}$ is the loss reduction in the offline user modeling task and $\Delta \mathrm{NE}_{\text{downstream}}$ is the realized gain in the online ranking task. This metric allows us to determine which scaling axes best preserve gains through the information bottleneck inherent in multi-stage architectures.

\section{Experimental Results}
\label{sec:experiments}

\subsection{Experimental Setup}
\label{subsec:scaling-setup}

\paragraph{Training infrastructure.}
All scaling experiments are conducted on the same large-scale production recommendation task using a proprietary industrial training platform. As previously noted, our experiments are conducted on the large upstream model to investigate the maximum achievable FLOP horizon, extending beyond the computational limits of the online model. Models are trained with mixed precision on \textbf{128} NVIDIA H100 GPUs and use FlashAttention~\cite{flash_attention} for memory and throughput efficiency. Unless otherwise noted, each configuration is trained for \textbf{229K} steps on \textbf{30 billion}  examples sampled from the production traffic distribution.

\paragraph{Evaluation metric.}
We report all results using \emph{Normalized Entropy} (NE), defined in Eq.~\ref{eq:normalized_entropy}. We report relative reductions in NE (\%) compared to a strong production baseline, where lower NE values indicate better performance. On our internal datasets, a reduction in NE of \textbf{0.02\%} is considered statistically significant and sufficient to yield measurable revenue impact.

\subsection{Scaling Model Depth and Width}
\label{subsec:scaling-depth-width}

We first investigate how to allocate a fixed compute budget between model depth ($L$) and width ($d$). We conduct a grid search over $L \in \{1,2,4,8\}$ and $d \in \{128,256,512,1024\}$ at a fixed sequence length of $T=400$.

The results reveal a non-linear relationship between depth and width. When the model is narrow ($d=128$), increasing depth yields diminishing returns; the representation capacity is bottlenecked by the embedding dimension. However, once width reaches a critical threshold of $d \approx 256$, depth scaling becomes substantially more effective. Conversely, allocating compute purely to width (fixing $L$ and increasing $d \to 1024$) yields limited NE gains.

To illustrate these regimes, Table~\ref{tab:config-comparison} compares four representative configurations. We include a ``Deep-Balanced'' variant to demonstrate the scaling behavior when both depth and width are sufficient. Extreme or unbalanced allocation to either width or depth leads to diminishing returns. Sufficient width ($d \ge 256$) is required for depth scaling to be effective, while excessive width without adequate depth (e.g., $d=1024$) is computationally inefficient and yields limited gains. Balanced configurations maximize performance and compute efficiency.

\begin{table}[h]
\centering
\caption{Capacity allocation across depth and width. \textbf{Deep-Balanced} ($d=256$) successfully leverages depth to achieve the best performance (-0.17\%)}
\label{tab:config-comparison}
\small
\begin{tabular}{lrrrr}
\toprule
\textbf{Setting} & \textbf{Deep} & \textbf{Balanced} & \textbf{Deep} & \textbf{Shallow} \\
 & \textbf{Narrow} & & \textbf{Balanced} & \textbf{Wide} \\
\midrule
Depth ($L$) & 8 & 4 & 8 & 2 \\
Width ($d$) & 128 & 256 & 256 & 1024 \\
\midrule
Seq Params (M) & 0.85 & 1.16 & 2.33 & 6.93 \\
Seq FLOPs (M)  & 568  & \textbf{530}  & 1187 & 1909 \\
\midrule
\textbf{NE $\Delta$ (\%)} & -0.10\% & -0.14\% & \textbf{-0.17\%} & -0.08\% \\
\bottomrule
\end{tabular}
\end{table}




Motivated by these observations, we adopt balanced configurations ($d \ge 256$) as the default when studying other scaling axes such as sequence length and content features.

\subsection{Sequence Length, Composition, and Content}
\label{subsec:sequence-quality}

We now examine how LLaTTE performance scales with the temporal horizon proportional to the number of events $T$, and with the semantic content embeddings on each event.

\subsubsection{Extending the temporal horizon}

\begin{figure}[t]
   \centering
   \includegraphics[width=0.95\columnwidth]{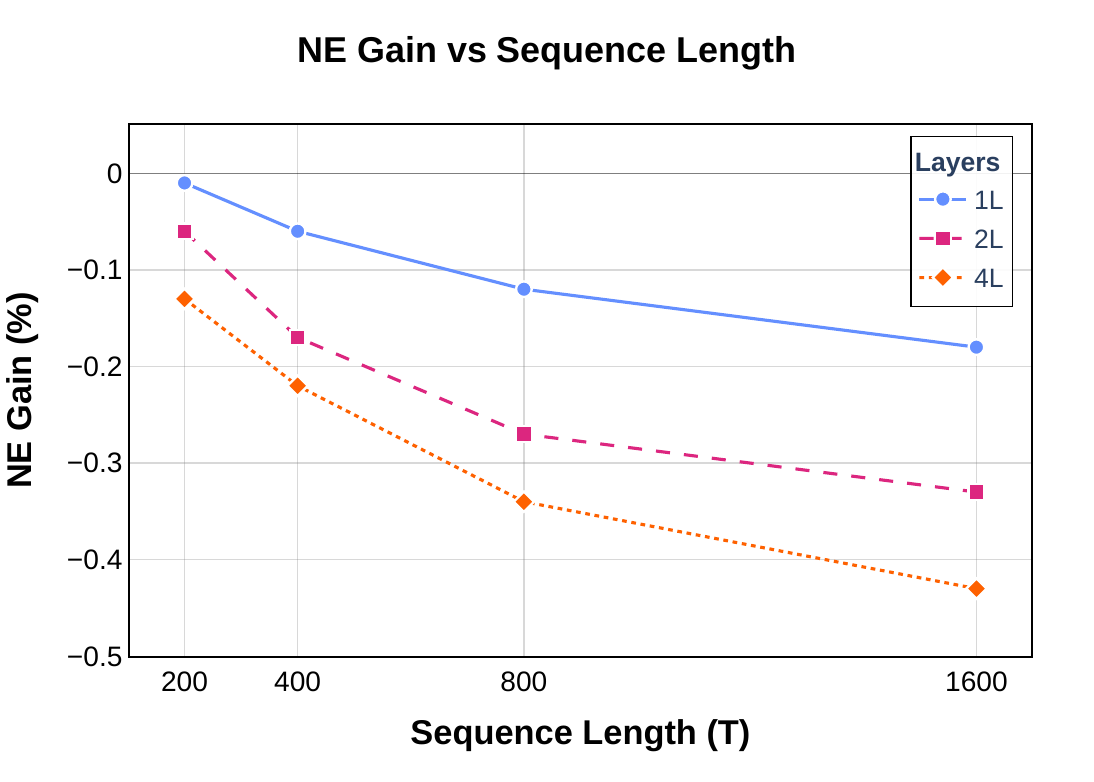}
   \caption{Sequence length scaling for $L=1$, $L=2$, and $L=4$ models. NE improves smoothly with longer histories, with deeper models exhibiting a steeper dependence on $T$.}
   \label{fig:seq_len_scaling}
\end{figure}

We vary the sequence horizon $T \in \{200, 400, 800, 1600\}$ for models with depths $L \in \{1,2,4,8\}$ while keeping width fixed. As summarized in Figure~\ref{fig:seq_len_scaling}, NE decreases monotonically with $T$ across all depths, confirming that longer user histories consistently improve ranking quality.

The incremental benefit of additional context depends strongly on model capacity. Deeper models exhibit larger NE gains as $T$ increases, which is reflected in the curves for $L=2$ compared to $L=1$ in Figure~\ref{fig:seq_len_scaling}. 

\begin{figure}[t]
\centering
\includegraphics[width=0.95\columnwidth]{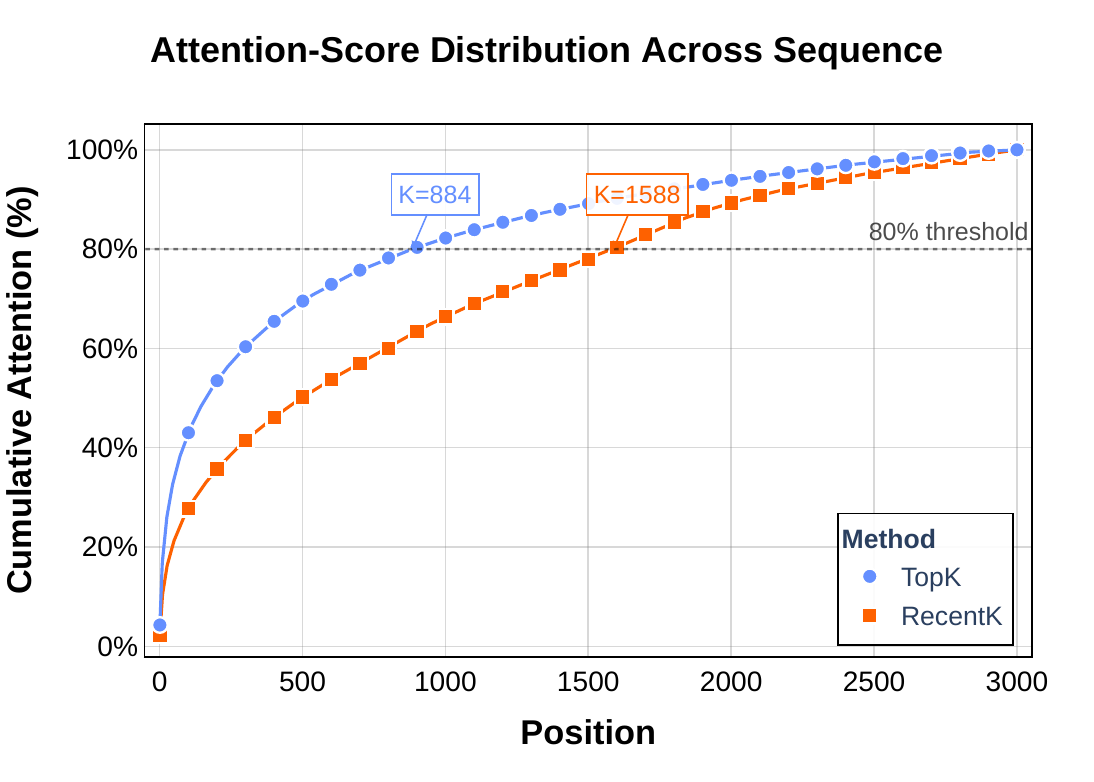}
\caption{Attention-score distribution across the sequence. The model continues to attend to events throughout the history, supporting the usefulness of long contexts.}
\label{fig:seq_len_attention}
\end{figure}

Attention-score probability analyses also revealed the utility of long contexts. The cumulative attention probability distributions over positions (Figure~\ref{fig:seq_len_attention}) show that a non-trivial fraction of attention probability is allocated across the entire $200$–$1600$ token range, rather than narrowly focusing on the most recent events, indicating that mid- and long-range history contributes meaningfully to predictions. The details of this study can be found in the Appendix (see Section~\ref{sec:Attention_weight}).

\subsubsection{Balancing freshness and signal strength}

We next examine the composition of the sequence at fixed length. In this setting, user histories combine high-frequency, lower per-event signal actions (ad views) with low-frequency, high-value actions (conversions). We fix $T=1000$ and vary the allocation between these two sources (Table~\ref{tab:sequence-quality}, top).

The balanced mixture of views and conversions performs best; both extremes are clearly suboptimal. Pure-view sequences underperform because they contain many low-signal events. Pure-conversion sequences, while composed of very high-value events, also degrade performance because conversions are temporally sparse: filling a length-$1000$ sequence with only conversions forces the model to rely on older events that may no longer reflect current intent. Views, by contrast, provide dense and recent coverage. The best-performing allocations combine the high per-event signal of conversions with the temporal freshness provided by frequent views.

\subsubsection{Content-aware scaling: strong content features bend the scaling curve}

Finally, we study how feature richness interacts with architectural scaling. In addition to sparse IDs, we introduce dense content embeddings produced by fine-tuned LLaMA models as well as Content Understanding models processing heterogeneous multimodal signals (e.g., text and images) and compare models with and without these features at different depths (Table~\ref{tab:sequence-quality}, bottom).

\begin{table}[h]
\centering
\caption{Sequence-quality ablations. Top: impact of sequence composition at fixed length $T=1000$. Bottom: effect of content semantic features at two depths.}
\label{tab:sequence-quality}
\small
\begin{tabular}{lcc}
\toprule
\multicolumn{3}{c}{\textbf{Sequence composition ($T=1000$)}} \\
\midrule
\textbf{Allocation (Views / Conv)} & \textbf{$\Delta$NE (\%)} &  \\
Balanced (500 / 500)   & \textbf{0.00\%}   & reference \\
Conv-heavy (200 / 800) & +0.01\%           & near-optimal \\
Pure conversions (0 / 1000) & +0.105\%     & degraded \\
Pure views (1000 / 0)  & +0.15\%           & worst \\
\midrule
\multicolumn{3}{c}{\textbf{Content features}} \\
\midrule
\textbf{Features}                 & \textbf{1L model} & \textbf{4L model} \\
Sparse IDs only                   & +0.06\%           & -0.01\% \\
Sparse IDs + content embeddings   & \textbf{0.00\%}   & \textbf{-0.118\%} \\
\bottomrule
\end{tabular}
\end{table}

Two observations stand out. First, in the absence of content features, increasing depth from one to four layers yields only a minor improvement: the 4-layer ID-only model is only slightly better than the 1-layer baseline. This corresponds to a very shallow scaling slope and indicates that additional capacity is largely spent memorizing ID patterns. Second, when LLaMA content embeddings are enabled, the same 4-layer configuration achieves a substantially larger gain, and the relative benefit of content is markedly stronger for the deeper model than for the shallow one.

These results show that strong semantic content features are not a marginal add-on but a prerequisite for effective scaling. The presence of semantic features materially changes the slope of the compute–NE curve: with ID-only inputs, depth and sequence-length scaling quickly exhibit diminishing returns, whereas with content-enriched sequences, the same increases in $L$ and $T$ translate into substantially larger NE gains. In this sense, the scaling law for LLaTTE is inherently \emph{content-aware}; analyzing model size and compute in isolation is insufficient to predict performance.

\subsection{Global Compute Scaling}
\label{subsec:global-scaling}

\begin{figure}[t]
\centering
\includegraphics[width=0.95\columnwidth]{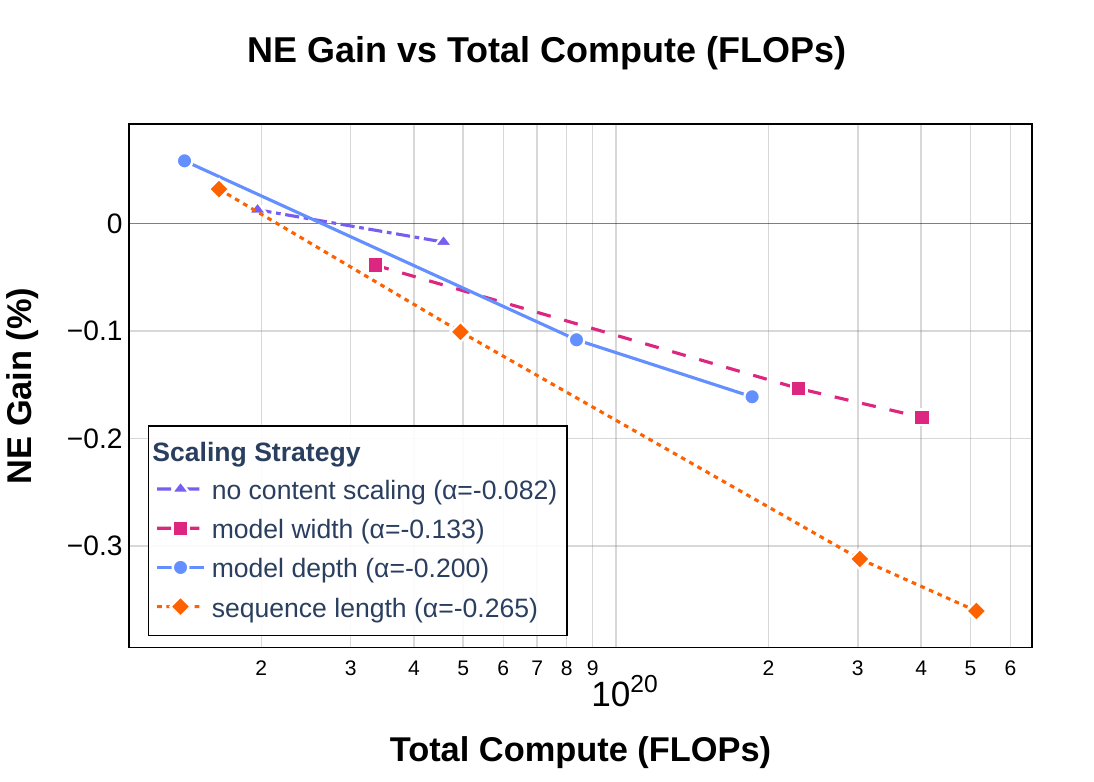}
\caption{Relative normalized entropy (NE) gain versus training compute (FLOPs, log scale) for different scaling strategies. The slope of each line indicates NE improvement per $10\times$ increase in compute.}
\label{fig:global_scaling}
\end{figure}

Finally, we aggregate the above experiments to examine global compute scaling. Figure~\ref{fig:global_scaling} plots relative NE gain versus sequence compute budget $\mathcal{C}$ (measured in sequence FLOPs) on a logarithmic scale.

The results demonstrate that recommendation performance follows a predictable power-law relationship with compute. We fit a log-linear model to the empirical data:

\[
\Delta \mathrm{NE}(\mathcal{C}) = - \alpha \cdot \log_{10} \mathcal{C} + \beta
\]

where the scaling coefficient $\alpha$ is determined by the scaling strategy. We observe distinct behaviors for each axis:

\begin{itemize}[leftmargin=*, noitemsep]

    \item \textbf{Sequence Length:} This dimension exhibits the steepest scaling slope. Extending the temporal horizon consistently yields the largest performance improvement per unit of compute, indicating that the model effectively utilizes long-term history. This, however, comes with the additional cost of logging and materializing longer user engagement histories.

    \item \textbf{Model Depth:} Increasing depth provides robust scaling behavior. Once the model width is sufficiently scaled up, adding layers offers a reliable path to improved ranking quality.

    \item \textbf{Model Width:} Width plays a foundational role in the architecture. While the scaling slope is less steep compared to depth or length, sufficient width is essential for maximizing the capacity of the other dimensions.

    \item \textbf{Content Quality:} As shown in Figure~\ref{fig:global_scaling}, the presence of semantic features significantly enhances the scaling laws on all other dimensions. Content-enriched models demonstrate a steep slope compared to their ID-only counterparts (black line), showing that high-quality semantic features are necessary for the model to extract value from increased compute.

\end{itemize}

These findings establish that sequence-based recommendation is a scalable paradigm that converts compute into predictable performance gains. We leverage this scaling hierarchy in Section~\ref{sec:multi-stage} to optimize our multi-stage architecture under production constraints.

\section{Multistage Sequence Model: Towards Maximizing RoI in Production}
\label{sec:multi-stage}

\subsection{Motivation and architecture}
\label{sec:multi-stage-motivation}

Section~\ref{subsec:global-scaling} showed that ranking quality continues to improve as we scale depth, and in particular, sequence length. The production ranker, however, serves trillions of requests per day under a strict latency budget, which limits the deployed sequence module to a few layers and a per-source (i.e. engagement type) horizon of roughly $T \approx 400$ events/tokens. To benefit from the favorable scaling curves without violating latency constraints, we adopt a multistage architecture. The latency-constrained \textbf{downstream} model remains compact and attends jointly over user sequences and ad/context features. An asynchronous \textbf{upstream} user model processes longer histories using the same LLaTTE architecture but with only user-side features, and publishes cached user embeddings that the ranker consumes at request time. This separation introduces a strict information bottleneck, as the upstream encoder must summarize thousands of historical events into a single fixed-size vector (in our production deployment, $d_{\text{transfer}} = 2048$). We study how this fixed-bandwidth bottleneck affects scaling laws.

\subsection{Upstream scaling laws}
\label{subsec:upstream-scaling}

We first characterize the scaling behavior of the upstream user model in isolation. Using the same log-linear fitting procedure as in Section~\ref{subsec:global-scaling}, we estimate slopes for NE improvement as a function of upstream sequence FLOPs. Table~\ref{tab:upstream-scaling} (top) summarizes the comparison between upstream and downstream slopes.

\begin{table}[h]
\centering
\caption{Upstream vs downstream scaling comparison.}
\label{tab:upstream-scaling}

\begin{subtable}{\linewidth}
\centering
\caption{Slopes $\alpha$ against total model FLOPs ($\mathcal{C}_{\text{full}}$).}
\label{tab:scaling-slopes-full}
\small
\begin{tabular}{lccc}
\toprule
                            & \textbf{Depth ($L$)} & \textbf{Width ($d$)} & \textbf{Seq.\ length ($T$)} \\
\midrule
\textbf{Downstream $\alpha$}  & 0.200                & 0.133                & 0.265 \\
\textbf{Upstream $\alpha$}    & 0.102                & 0.113                & 0.116 \\
\bottomrule
\end{tabular}
\end{subtable}

\vspace{1em}

\begin{subtable}{\linewidth}
\centering
\caption{Slopes $\alpha$ against sequence-only FLOPs ($\mathcal{C}_{\text{seq}}$).}
\label{tab:scaling-slopes-seq}
\small
\begin{tabular}{lccc}
\toprule
                            & \textbf{Depth ($L$)} & \textbf{Width ($d$)} & \textbf{Seq.\ length ($T$)} \\
\midrule
\textbf{Downstream $\alpha$}  & 0.106                & 0.091                & 0.238 \\
\textbf{Upstream $\alpha$}    & 0.092                & 0.102                & 0.094 \\
\bottomrule
\end{tabular}
\end{subtable}

\vspace{1em}

\begin{subtable}{\linewidth}
\centering
\caption{Transfer ratios for iso-FLOPs upstream configurations.}
\label{tab:iso-flops-transfer}
\small
\begin{tabular}{lcc}
\toprule
                            & \textbf{Seq-heavy}              & \textbf{Model-heavy}             \\
                            & ($L{=}3, d{=}512, T{=}1000$)    & ($L{=}6, d{=}512, T{=}460$)      \\
\midrule
$\Delta$NE$_{\text{up}}$      & $-0.14\%$                       & $-0.13\%$                        \\
$\Delta$NE$_{\text{down}}$    & $-0.07\%$                       & $-0.07\%$                        \\
$\tau$ (\%)                   & 50                              & 53                               \\
\bottomrule
\end{tabular}
\end{subtable}

\end{table}

Table~\ref{tab:scaling-slopes-full} reports scaling slopes against total model FLOPs. The downstream model exhibits notably steeper slopes across all dimensions. This difference, however, reflects a structural asymmetry: in the downstream ranker, the sequence module accounts for only $\approx 30\%$ of total FLOPs, whereas in the upstream model it accounts for $\approx 90\%$. Scaling the sequence module thus yields a larger relative improvement per total FLOP in the downstream setting.

To isolate the intrinsic scaling efficiency of the sequence module, Table~\ref{tab:scaling-slopes-seq} reports slopes against sequence-only FLOPs. This comparison reveals that depth and width scaling efficiency is consistent across both settings. The key difference lies in sequence length: upstream scaling retains only $\approx 50\%$ of the downstream efficiency. This gap arises because the upstream encoder lacks candidate context, forcing it to compress the history into a generic representation rather than performing candidate-aware attention. Consequently, only long-range signals that are universally relevant across candidates survive the bottleneck.

\subsection{Transfer efficiency}
\label{subsec:transfer-ratio}

To quantify the efficiency of our multi-stage architecture, recall the transfer ratio $\tau$ defined in Eq.~\ref{eq:transfer_ratio}. This metric measures the conversion rate of upstream representation quality into downstream ranking accuracy.



For instance, the \textit{Seq-heavy} configuration in Table~\ref{tab:iso-flops-transfer} demonstrates this relationship: upgrading the upstream encoder yields a $0.14\%$ upstream improvement, which translates to a $0.07\%$ improvement in the downstream ranker, corresponding to a transfer ratio of $\tau \approx 50\%$.



\paragraph{Benchmarking Transfer Efficiency.}
Transferring gains from asynchronous upstream models to online rankers is inherently lossy. Several structural factors typically limit transfer efficiency to the $25\%\text{--}30\%$ range: the \textit{capacity gap} between massive upstream encoders and compact downstream rankers, the \textit{staleness} introduced by asynchronous inference, and the \textit{information bottleneck} imposed by compressing user histories into fixed-size user embeddings. Against this backdrop, the $\approx 50\%$ transfer ratio observed in our experiments is significant. It indicates that the high-level intent signals captured by scaling the sequence encoder are robust to compression and temporal delays, retaining substantial predictive value despite the aggressive compression and separation from the online scoring context.

\paragraph{Architecture Robustness.}
To determine if specific architectural choices affect this efficiency, we compare two upstream configurations with matched compute budgets ($\approx 12$ GFLOPs/sample) representing different scaling priorities:

\begin{itemize}[leftmargin=*, noitemsep]
    \item \textbf{Sequence-heavy}: $L=3, d=512, T=1000$
    \item \textbf{Model-heavy}: $L=6, d=512, T=460$
\end{itemize}

As shown in Table~\ref{tab:upstream-scaling} (bottom), both models achieve similar upstream gains ($-0.14\%$ vs $-0.13\%$) and, crucially, identical downstream gains ($-0.07\%$). This indicates that for a fixed compute budget, the architecture is robust to the specific allocation between depth and sequence length. The total compute of the upstream model determines the downstream performance, validating that we can flexibly trade off history length for model capacity without degrading the transfer ratio.

\subsection{Implications for compute allocation}
\label{subsec:compute-allocation}

The combined downstream and upstream observations yield a simple policy:

\begin{itemize}[leftmargin=*,noitemsep]
    \item \textbf{Downstream online ranker:} Allocate the latency budget to short sequence lengths and shallow depths, supported by sufficient width and rich content features (sequence lengths up to $T \approx 400$).
    \item \textbf{Asynchronous upstream user model:} Use the relaxed latency budget to scale total sequence-modeling compute. Results indicate that downstream performance is driven primarily by the total upstream compute budget rather than the specific allocation between depth and length. Consequently, we deploy upstream models with a sequence compute budget $45\times$ larger than that of the online ranker.
\end{itemize}

This two-stage asymmetric strategy lets us continue following the scaling frontier identified in Section~\ref{subsec:global-scaling} after the latency-constrained ranker has exhausted its local budget, turning additional offline compute into predictable online gains.
\section{Production Deployment}
\label{sec:prod_dep}

\subsection{Asynchronous embedding service with online LLaTTE}

In production, we deploy a two-stage LLaTTE architecture. The \emph{downstream} ranking model continues to run a compact LLaTTE module online, attending over short user histories (per-source horizon capped at $T \approx 400$) together with ad and context features, under a strict per-request latency budget at trillion-request scale. This online LLaTTE model captures fresh user intent signals and leverages ad-specific interactions, but is intentionally much smaller than the models investigated in the scaling study presented in Section~\ref{subsec:global-scaling}.

To exploit bigger sequence models without impacting latency, we introduce an \emph{asynchronous embedding service} that hosts the \emph{upstream} user understanding LLaTTE model. Embedding updates are not computed per request; instead, they are triggered on high-value user events (primarily user conversions). Triggered events are processed on a dedicated cluster of H100 GPU hosts using deeper and longer LLaTTE models optimized for high-throughput transformer inference. Representing the largest user model deployment at Meta, these upstream encoders operate only on user-side features, generate user representations (embeddings), and write their compressed representations to a feature store. At request time, the downstream ranking models read these embeddings as dense features and utilize them in the online downstream LLaTTE ranking model.


\subsection{Latency and online gains}

Offloading heavy sequence modeling to the asynchronous embedding service keeps the additional cost on the ranking path to a single feature lookup. The online downstream LLaTTE model remains conservative in design (depth and sequence length). We observe no measurable change in P99 ranking latency compared to the baseline without upstream LLaTTE user embeddings. The extra FLOPs from the larger upstream LLaTTE models are absorbed by the H100 cluster, which operates at much lower QPS and is heavily batched.

Across multiple large-scale A/B tests, the combination of the compact online LLaTTE and the upstream LLaTTE user embeddings delivered approximately \textbf{0.25\%} NE reduction on our flagship ads ranker, corresponding to a 4.3\% conversion uplift on Facebook Feed and Reels, a business impact on the order of hundreds of millions of dollars in annual revenue. This validates the proposed two-stage scaling strategy as an efficient way to convert additional sequence-modeling compute into production value under strict serving constraints.

\section{Conclusion \& Future Work}
\label{sec:conclusion}

In this work, we propose the LLaTTE paradigm, which significantly scales up sequence learning in recommender systems. By allocating the majority of computational resources to sequence modeling, we achieve predictable and robust scaling behavior. We systematically analyze scaling across standard dimensions (model depth and width) as well as less explored factors (sequence length, composition and richness), demonstrating that optimal scaling requires balancing all these factors simultaneously.

To maximize production impact, we develop a multistage modeling and deployment strategy: a computationally expensive model trained on user-only features that runs offline and infrequently, while effectively transferring these gains to the lightweight online production model.

Our work was developed concurrently with other notable industry advances in sequence modeling for recommender systems~\cite{onetrans2025, onerec}. These efforts collectively demonstrate the rise of sequence modeling in large-scale recommendation systems. Moving forward, 
we plan to explore efficient long-context kernels, reinforcement learning, scalable infrastructure solutions, and the upper bounds of scaling laws -- as we stand at the cusp of the LLM-scale era for recommender systems.

\bibliographystyle{ACM-Reference-Format}
\bibliography{references}

\appendix

\section{LLaTTE Architecture Details}
\label{app:llatte-arch}

\subsection{Non-Sequence Module}

Let $\mathcal{E}: \mathcal{V} \to \mathbb{R}^{d_e}$ denote the embedding lookup for sparse categorical features. We construct the initial non-sequence representation by concatenating embedded sparse features, dense features, float features, and sequence summaries:
\begin{equation}
\mathbf{h}^{(0)} = \text{Concat}\Big(
\mathcal{E}(\mathbf{x}_{\text{sparse}}),\;
\mathbf{x}_{\text{dense}},\;
\mathbf{x}_{\text{float}},\;
\{\mathbf{z}^{(k)}_{\text{seq}}\}_{k=1}^{m_{\text{seq}}}
\Big) \in \mathbb{R}^{d_0}.
\end{equation}
We then apply $L_{\text{NS}}$ layers of a feature-interaction network:
\begin{equation}
\mathbf{h}^{(\ell)} = \text{NonSeq}_\ell(\mathbf{h}^{(\ell-1)}), \quad \ell = 1, \dots, L_{\text{NS}},
\end{equation}
yielding
\begin{equation}
\mathbf{z} = \mathbf{h}^{(L_{\text{NS}})} \in \mathbb{R}^{d}.
\end{equation}
In production, $\text{NonSeq}_\ell$ is instantiated with a DCN/DeepFM/DLRM-style architecture~\cite{guo2017deepfm, naumov2019dlrm, wang2017dcn, wang2021dcnv2, wukong2024}.

\subsection{Sequence Module}

\subsubsection{Action Embeddings}

Let $S_u = \{a_1, a_2, \ldots, a_T\}$ be the temporally ordered sequence of actions, oldest first. Each action
$a_t = (\tau_t, \text{type}_t, \text{item}_t, \text{surface}_t, \text{meta}_t)$
is embedded and projected into a token:
\begin{align*}
\mathbf{x}_t = \text{MLP}_{\text{act}}\Big(
    &\mathcal{E}_{\text{type}}(\text{type}_t),\;
    \mathcal{E}_{\text{item}}(\text{item}_t), \\
    &\mathcal{E}_{\text{surface}}(\text{surface}_t),\;
    \mathcal{E}_{\text{time}}(\tau_t),\;
    \mathcal{E}_{\text{meta}}(\text{meta}_t)
\Big) \in \mathbb{R}^{d}
\end{align*}
forming
\begin{equation}
\mathbf{X}_{\text{seq}} = [\mathbf{x}_1; \mathbf{x}_2; \ldots; \mathbf{x}_T] \in \mathbb{R}^{T \times d}.
\end{equation}
We apply additive timestamp encodings to a subset of hidden dimensions.

\subsubsection{Query Tokens and Fusion}

We introduce $n_q$ query tokens $\mathbf{Q} \in \mathbb{R}^{n_q \times d}$ that summarize the user–ad request. Depending on the stage, these tokens may encode candidate ad features, request context, user-level features, and learned seed tokens~\cite{jaegle2021perceiver}. We concatenate sequence and query tokens:
\begin{equation}
\mathbf{X}_{\text{input}} = \text{Concat}(\mathbf{X}_{\text{seq}}, \mathbf{Q}) \in \mathbb{R}^{(T + n_q) \times d},
\end{equation}
and feed $\mathbf{X}_{\text{input}}$ into the transformer.

\subsubsection{Transformer Layers and Pyramidal Schedule}

We apply an $L$-layer causal transformer with Multi-head Latent Attention (MLA)~\cite{deepseek2024mla} and RMS normalization~\cite{zhang2019rootmeansquarelayer}. Let $\mathbf{R}^{(0)} = \mathbf{X}_{\text{input}}$. For layer $\ell$:
\begin{align}
    \mathbf{Z}^{(\ell)} &= \text{RMSNorm}\big(\mathbf{R}^{(\ell)} + \text{MLA}(\mathbf{R}^{(\ell)})\big), \\
    \mathbf{X}^{(\ell)} &= \text{RMSNorm}\big(\mathbf{Z}^{(\ell)} + \text{FFN}(\mathbf{Z}^{(\ell)})\big).
\end{align}

Following~\cite{onetrans2025}, we use an adaptive pyramidal schedule over the temporal dimension. If $\mathbf{X}^{(\ell)} \in \mathbb{R}^{(T_\ell + n_q) \times d}$, the next layer operates on
\begin{equation}
\mathbf{R}^{(\ell+1)} = \mathbf{X}^{(\ell)}[:, -T_{\ell+1}:, :] \in \mathbb{R}^{T_{\ell+1} \times d},
\end{equation}
where $T_{\ell+1} \leq T_\ell$. This progressively trims older tokens, reducing attention cost $\mathcal{O}(T_{\ell+1} T_\ell d)$ and FFN cost $\mathcal{O}(T_{\ell+1} d^2)$.

We use three regimes:
\begin{itemize}[leftmargin=*,noitemsep]
    \item \textbf{Full self-attention} ($T_\ell = T + n_q$): all sequence and query tokens are retained.
    \item \textbf{Pyramidal attention} ($n_q < T_\ell < T + n_q$): query tokens plus the $m = T_\ell - n_q$ most recent actions.
    \item \textbf{Cross-attention} ($T_\ell = n_q$): only query tokens are retained; used in the final layer.
\end{itemize}

Offline (upstream) models typically use more layers with full self-attention (except the final cross-attention layer). Online ranking models use more aggressive pyramidal trimming in early layers before the final cross-attention layer.

\subsubsection{Sequence Summaries}

The final layer outputs $\mathbf{Z}_{\text{raw}}$ at the query-token positions. We project these into one or more fixed-size sequence summaries:
\begin{equation}
\mathbf{z}^{(k)}_{\text{seq}} = \text{LoRAMLP}_k\big(\text{Flatten}(\mathbf{Z}_{\text{raw}})\big) \in \mathbf{R}^{d_{\text{seq}}},
\end{equation}
where each $\text{LoRAMLP}_k$ is a low-rank adapted MLP.

\subsection{Multi-head Latent Attention}
\label{sec:MLA_details}

For completeness, we include the MLA parameterization. For $h$ heads, given input $\mathbf{X} \in \mathbf{R}^{h \times d}$:
\begin{align}
    \mathbf{K}, \mathbf{V} \in \mathbb{R}^{T \times h \times d_k} 
    &= \text{split}\left(\text{RMSNorm}(\mathbf{X} \mathbf{W}^{KV}_{\text{down}}) \mathbf{W}^{KV}_{\text{up}}\right) \\
    \mathbf{Q} \in \mathbb{R}^{T \times h \times d_k} 
    &= \text{RMSNorm}(\mathbf{X} \mathbf{W}^{Q}_{\text{down}}) \mathbf{W}^{Q}_{\text{up}} \\
    \text{MLA}(\mathbf{X}) \in \mathbb{R}^{T \times d} 
    &= \text{softmax}\left(\frac{\mathbf{Q} \mathbf{K}^\top}{\sqrt{d_k}}\right) \mathbf{V} \mathbf{W}_{\text{out}}
\end{align}

A key observation is that consecutive linear projections can be algebraically absorbed 
into single matrices. Specifically:
\begin{itemize}
    \item the $\mathbf{W}^{Q}_{\text{up}}$ and the ${\mathbf{W}^{KV}_{\text{up}}}^\top$ on $\mathbf{K}^\top$ can be fused
    \item the $\mathbf{W}^{KV}_{\text{up}}$ on $Q$ can be absorbed into $\mathbf{W}_{\text{out}}$
\end{itemize}
This simplification reveals that MLA is mathematically equivalent to 
Multi-Query Attention (MQA)~\cite{shazeer2019fasttransformerdecodingwritehead} on the latent space, 
where queries are projected to multiple heads while key-value representations remain 
shared in the compressed latent space. The reformulated equations are as follows:

\begin{align}
    \mathbf{Q}_{\text{latent}} \in \mathbb{R}^{T \times h \times d_c} 
        &= \mathrm{RMSNorm}(\mathbf{X} \mathbf{W}^{Q}_{\text{down}}) \\
    \mathbf{KV}_{\text{latent}} \in \mathbb{R}^{T \times d_c} 
        &= \mathrm{RMSNorm}(\mathbf{X} \mathbf{W}^{KV}_{\text{down}}) \\
    \mathrm{MLA}(\mathbf{X}) \in \mathbb{R}^{T \times d} 
        &= \mathrm{softmax}\left(
            \frac{\mathbf{Q}_{\text{latent}} \mathbf{W}^{QK} \mathbf{KV}_{\text{latent}}^\top}{\sqrt{d_c}}
        \right) \nonumber \\
        &\quad \cdot \mathbf{KV}_{\text{latent}} \mathbf{W}^{V}_{\text{out}}
\end{align}

\subsection{Attention Weight Distribution}
\label{sec:Attention_weight}

We studied the attention weight distribution $P$
\begin{equation}
P =\text{softmax}\left(\frac{\mathbf{Q} \mathbf{K}^\top}{\sqrt{d_k}}\right)
\end{equation}

To enhance the visualization of attention weights, we reduced both the number of query tokens and the number of attention heads in a single-layer transformer LLaTTE backbone. We extracted the attention probabilities of each query token with respect to the entire user sequence and aggregated these probabilities across a sufficient number of samples from diverse user requests.

Our analysis reveals that the model allocates a greater proportion of its attention to the most recent events. However, it continues to assign non-negligible probabilities to longer-term history, indicating that the model does not disregard earlier events entirely Figure~\ref{fig:seq_len_attention}. Interestingly, we also examined the cumulative attention weight by selecting topK event tokens instead most recent k tokens, which present the theoretical maximum cumulative attention weight achievable by selecting k tokens.

\begin{figure}[t]
\centering
\includegraphics[width=0.8\columnwidth]{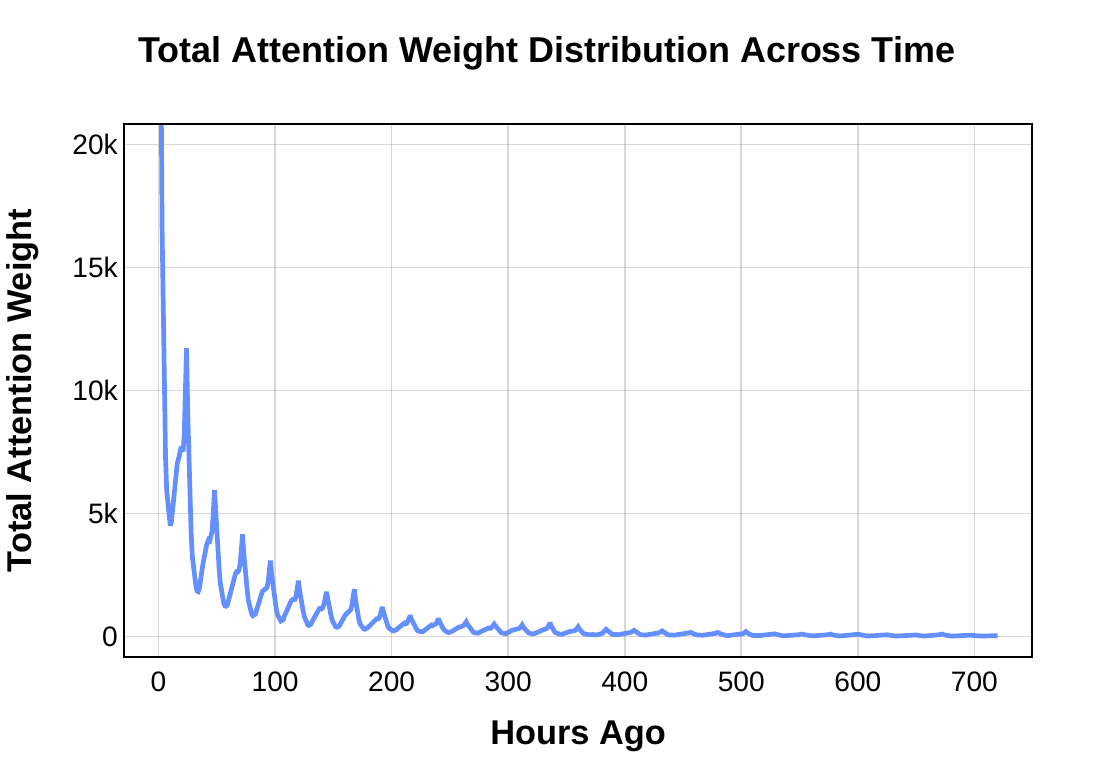}
\caption{Total attention weight distributed to event tokens bucketized by hours prior to the user request.}
\label{fig:hourly_total_attn}
\end{figure}

\begin{figure}[t]
\centering
\includegraphics[width=0.8\columnwidth]{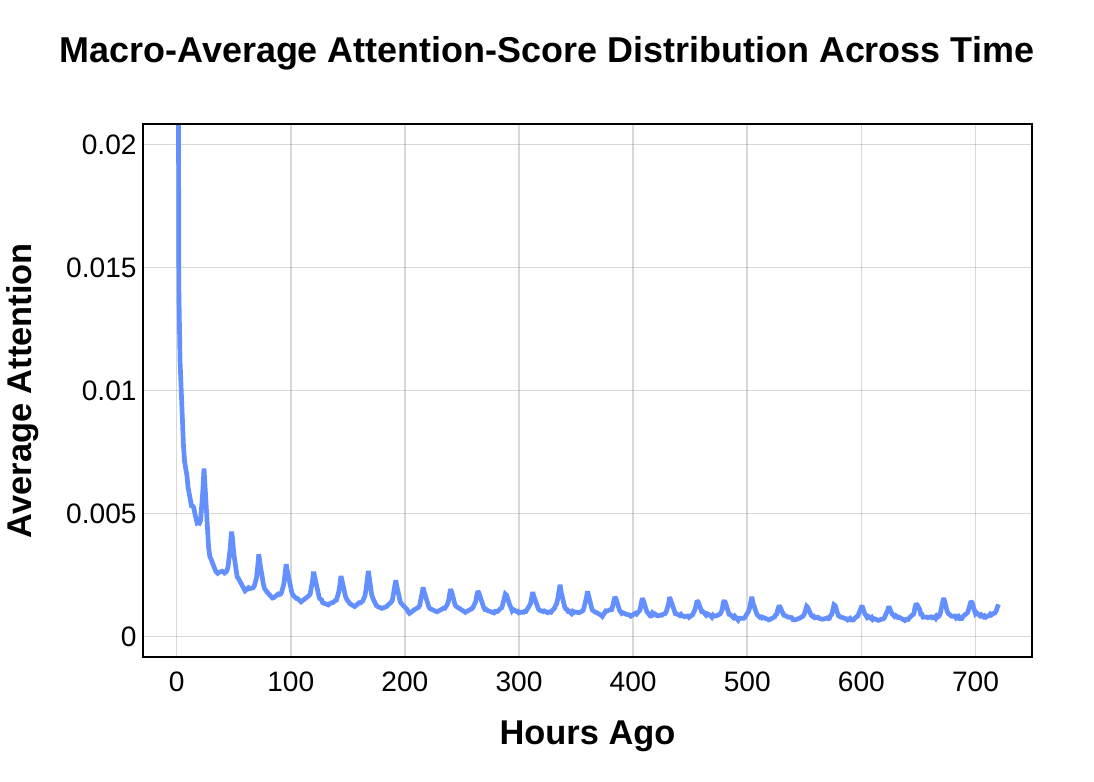}
\caption{Average attention weight per event by hours prior to the user request.}
\label{fig:avg_hourly_total_attn}
\end{figure}

Additionally, we identified notable seasonality patterns at the daily level. As shown in Figure~\ref{fig:hourly_total_attn} and Figure~\ref{fig:avg_hourly_total_attn}, the attention weight spikes occur at sharp 24-hour intervals. This suggests that individual users tend to exhibit recurring interests and behaviors at similar times of day.

While these findings are consistent with our intuitions, we recognize that a more granular understanding of user behavior in relation to behavioral sequences remains an open area for exploration. We present these preliminary results to encourage further research and contributions from the broader community.
\section{Acknowledgments}

We thank Rui Yang, Xinyi Zhao, Minji Wu, Zhaoyang Huang, Zhehui Zhou, Larry Zhang, Wenbo Bu, Liang Tao, Alex Li, Guang Yang, Ketan Singh, Dianshi Li, Ruichao Xiao, Dan Barysevich, Zahra Rezapour, Abdul Zainul-Abedin, Gaoxiang Liu, Rupert Wu, Yuxi Hu, Leo Ding, and Qiao Yang for infrastructure support.

We thank Zhenyuan Liu, Fan Xia, I-Ta Lee, Feng Wei, Michael Du, Bella Shi, Bella Zhang, Zhaoheng Zheng, Nabil Hossain, Nara Lakamsani, Mehrnoosh Mirtaheri, Samarth Mittal and Minghai Chen for developing strong content features.  

We are grateful to Julia Ulziisaikhan, Yang Zhang, Jingjing Feng, and Yuan Zhang for data science support, and to Abha Jain and Sean O'Byrne for product management guidance. We thank Danil Kirsanov, Rostam Shirani and Steven De Gryze for management support. 

We thank Mingwei Tang, Allen Lin, Xinlong Liu, Mustafa Acar, Reazul Russel, Navid Madani, Alan Yang, Ashish Katiyar, Metarya Ruparel, Kai Wang, Robert Tang, Eley Ng, James Teng, Manpreet Singh Takkar for productionization efforts.

\end{document}